# Solar Wind Proton Implantation and Hydrogen Diffusion: Model Comparisons to $M^3$ Observations

O. J. Tucker[1], W. M. Farrell[2], S. Li[3], A. R. Poppe[4], Kennedy White[5], Valeriy Tenishev[6]

[1]Hampton University, 154 William R. Harvey Way, Hampton Va, USA.

[2]Space Science Institute, 4765 Walnut Street, Suite B, Boulder, CO 80301, USA.

[3]Hawaii Institute of Geophysics and Planetology, University of Hawaii, Honolulu, HI, USA.

[4]Space Sciences Laboratory, University of California at Berkeley, Berkeley, CA, USA.

[5]University of California, Santa Cruz, Santa Cruz, CA, USA.

[6]NASA's Marshall Space Flight Center 4200 Rideout Road, Redstone Arsenal, AL 35812, USA.

**Acknowledgments:** OJT was supported by the NASA Solar System Exploration Research Virtual Institute (SSERVI) Lunar Environment and Dynamics for Exploration Research (LEADER) team (Grant No. 80NSSC20M0060). OJT and VT were supported by NASA's the Lunar Data Analysis Program (LDAP). ARP was supported by the THEMIS-ARTEMIS mission, grant No. NAS5-02099. SL and WMF were supported by the NASA SSERVI, The Center for Lunar Environment and Volatile Exploration Research (CLEVER) (80NSSC23M022).

# Solar Wind Proton Implantation and Hydrogen Diffusion: Model Comparisons to $M^3$ Observations

O. J. Tucker[1], W. M. Farrell[2], S. Li[3], A. R. Poppe[4], Kennedy White[5], Valeriy Tenishev[6]

[1]Hampton University, 154 William R. Harvey Way, Hampton Va, USA.

[2]Space Science Institute, 4765 Walnut Street, Suite B, Boulder, CO 80301, USA.

[3]Hawaii Institute of Geophysics and Planetology, University of Hawaii, Honolulu, HI, USA.

[4]Space Sciences Laboratory, University of California at Berkeley, Berkeley, CA, USA.

[5]University of California, Santa Cruz, Santa Cruz, CA, USA.

[6]NASA's Marshall Space Flight Center 4200 Rideout Road, Redstone Arsenal, AL 35812, USA.

## Abstract

The Moon Mineralogy Mapper ($M^3$) observed a widespread 3-micron absorption feature on the lunar surface, commonly attributed to surficial OH and/or $H_2O$. The feature varies with latitude, local time, and lunar phase, suggesting that at least part of the optically active reservoir is dynamic. We compare phase- and local-time-resolved $M^3$ abundance estimates with a global solar wind proton implantation, hydrogen diffusion, and $H_2$ exosphere model. The model tracks implanted H in the upper grain rim, thermally activated retention and loss, recombinative $H_2$ release, and phase-driven source variations associated with the solar wind, magnetosheath, and magnetotail. We compute the surface abundance using a weighted activation-energy-bin representation, in which the activation-energy distribution is discretized into fixed probability-weighted bins rather than sampled stochastically by Monte Carlo particles. This implementation reduces particle sampling noise in low-flux regions. A refined grid of Gaussian effective activation-energy distributions shows that the low- and mid-latitude $M^3$ trends are best matched by distributions centered near $E_c = 0.520$ eV with width $E_w = 0.090$ eV. The preferred case gives a combined low- and mid-latitude RMSE of approximately 27 ppm and a small mean bias of only a few ppm. These fitted parameters should be interpreted as an effective response of the $M^3$-sensitive retained H/OH reservoir, not as a unique mineralogical activation energy. The model reproduces the dominant local-time and phase-dependent abundance structure, including the reduced retained abundance near local noon, while model-data differences near plasma-transition intervals and high-latitude behavior may reflect additional sampling, saturation, or surface-plasma processes.

## Plain Language Summary

Hydrogen from the solar wind can strike the Moon and become implanted in the uppermost layer of lunar soil grains. Some of this hydrogen can attach to oxygen-bearing sites in the grains, producing hydroxyl-like bonds that contribute to the 3-micron signal measured by the Moon Mineralogy Mapper ($M^3$). Because the lunar surface heats and cools each day, implanted hydrogen can be retained more efficiently in cooler regions and lost more rapidly from warmer regions. We tested whether this process explains the $M^3$ measurements by comparing the observations with a global model of solar wind implantation, hydrogen diffusion, and $H_2$ release from the lunar surface. The model was sampled at the same lunar phase, local time, and latitude bins as the $M^3$ data. The comparison suggests that the lower abundance observed near local noon can be explained largely by temperature-driven hydrogen loss from the surface. We find that the low- and mid-latitude observations are reproduced well when the model uses an effective range of diffusion activation energies centered near 0.52 eV. This value should not be interpreted as one specific mineral or defect-site activation energy. Instead, it describes the portion of implanted hydrogen that is active over lunar day-to-month timescales. The remaining differences between model and data may point to additional effects such as plasma interactions, surface heterogeneity, or high-latitude saturation.

## 1. Introduction

A major lunar discovery reported in 2009 was the detection of a near 3-micron infrared absorption feature at the lunar surface by Chandrayaan-1's Moon Mineralogy Mapper ($M^3$) (Pieters et al., 2009), Cassini's Visual and Infrared Mapping Spectrometer (VIMS) (Clark, 2009), and the Deep Impact High Resolution Instrument-Infrared spectrometer (HRI-IR) (Sunshine et al., 2009). This absorption feature is generally interpreted as evidence for hydroxyl and/or molecular water in the optically active upper surface of lunar regolith grains. The observation was unexpected because the Moon had long been considered largely anhydrous at its surface. The reported increase in absorption strength toward higher latitudes and its time-of-day dependence suggested that at least part of the surficial OH/$H_2O$ reservoir is dynamic rather than purely indigenous or static.

Solar wind proton implantation has been widely considered as a primary process contributing to surficial lunar hydroxylation. In this pathway, incident solar wind protons are implanted into the upper tens of nanometers of regolith grains, become neutralized, and interact with oxygen-rich lattice sites to form metastable O-H bonds. The resulting OH-bearing population can then be modified by temperature-dependent diffusion, recombination, and release to the exosphere. This interpretation has motivated observational, experimental, and modeling studies of solar-wind-driven lunar hydroxylation, including laboratory irradiation studies of OH formation and $H_2$/$H_2O$ release, kinetic models of solar-wind-induced water formation, statistical-mechanical treatments of implanted-H retention, and global implantation-diffusion-exosphere models that couple retained surface H/OH to exospheric $H_2$ release (McCord et al., 2011; Klima et al., 2013; Farrell et al., 2015, 2017; Li and Milliken, 2017; Woehler et al., 2017; Bandfield et al., 2018; Jones et al., 2018; Tucker et al., 2019, 2021; Li et al., 2023; Jones et al., 2024; Lu et al., 2024).

Recently, Li et al. (2023) produced a high-level $M^3$ data set describing the inferred lunar surface water content as a function of local time, latitude, and lunar phase during periods when the Moon passed through the terrestrial magnetosphere. They found that the inferred $OH/H_2O$ abundance is elevated in the local morning and afternoon relative to local noon across broad latitude bands. This trend is consistent with temperature-dependent diffusive loss, in which warmer near-noon surface temperatures reduce the retained H/OH abundance. Li et al. (2023) also reported that the inferred surficial water content at comparable local times can remain similar between periods of direct solar wind exposure and periods when the Moon is inside Earth's magnetotail, despite the large reduction in incident solar wind proton flux.

This finding motivates a direct comparison between the $M^3$-derived $OH/H_2O$ trends and a global solar wind implantation, hydrogen diffusion, and $H_2$ exosphere model. Previous work by Tucker et al. (2019, 2021) showed that solar wind implanted H, diffusive retention in regolith grains, and recombinative release of $H_2$ can jointly explain key aspects of the lunar surface $OH/H_2O$ distribution and the lunar $H_2$ exosphere. However, the Li et al. (2023) data set enables a more direct comparison because it provides phase- and local-time-resolved constraints across the magnetotail, magnetosheath, and solar wind environments.

Here we re-examine the Tucker et al. (2019, 2021) model using the Li et al. (2023) phase- and local-time-resolved $M^3$ data set. We focus on whether the observed low- and mid-latitude $OH/H_2O$ trends can be reproduced by a solar wind implantation and thermally controlled diffusion model using an effective distribution of H diffusion activation energies. We compare a grid of Gaussian activation-energy distributions to the $M^3$ data at the corresponding lunar phase, local time, and latitude bins. This approach allows us to test whether the observations constrain a preferred effective activation-energy response range for the $M^3$-sensitive reservoir.

In this paper, Section 2 reviews the solar wind implantation and hydrogen diffusion pathway. Section 3 summarizes the $M^3$ data set and the phase/local-time sampling used for comparison. Section 4 describes the global surface diffusion-exosphere model and the activation-energy distribution used to represent implanted H retention. Section 5 describes the model-data comparison and parameter-space search used to constrain the effective activation-energy distribution. Section 6 presents the model results, including the activation-energy grid search, the preferred low- and mid-latitude comparison, and diagnostic residual structure across phase and local time.

## 2. Solar Wind Implantation and Hydrogen Diffusion

The idea that solar wind protons can produce OH or $H_2O$ in silica-like lunar regolith is not new. Early studies recognized that implanted hydrogen could interact with oxygen-bearing materials at the lunar surface (Zeller et al., 1966; Housley et al., 1973, 1974). Starukhina (2001, 2006) developed a physical picture in which solar wind protons are implanted into the upper damaged rims of lunar grains, diffuse through an oxygen-rich amorphous lattice, become temporarily trapped at defect sites, and are ultimately released from the surface, in part as molecular hydrogen. This process is illustrated schematically in Figure 1.

In this pathway, incident protons are implanted in a broad depth distribution centered near approximately 20 nm, consistent with ion-implantation calculations for solar-wind energies and the TRIM-based implantation depths adopted in Farrell et al. (2015, 2017) and Tucker et al. (2019, 2021). Once implanted, hydrogen atoms migrate through the damaged amorphous grain rim by thermally activated hopping among oxygen-related trapping sites, analogous to the hindered H diffusion described for irradiated silica and to molecular-dynamics simulations of H migration in amorphous silicates (Fink et al., 1995; Starukhina, 2006; Farrell et al., 2017; Morrissey et al., 2022). During migration, H atoms may form metastable O-H bonds. Because these bonds persist much longer than the vibrational period associated with the 3-micron absorption feature but can still be modified over diurnal or lunation timescales, the metastable O-H population is a plausible contributor to the remotely observed near-infrared-sensitive OH/$H_2O$ reservoir sampled by $M^3$, Cassini VIMS, and Deep Impact (Pieters et al., 2009; Clark, 2009; Sunshine et al., 2009).

Hydrogen diffusion in irradiated silica and related materials is commonly described using an Arrhenius-type diffusion coefficient,

$$D = D_0 \exp(-E/k_b T). \qquad \text{Eq. 1}$$

Eq. 1 defines the Arrhenius diffusion coefficient used to describe thermally activated H transport, where $D_0$ is the diffusion prefactor, $E$ is the activation energy, $k_b$ is Boltzmann's constant, and $T$ is temperature. Laboratory studies of hydrogen diffusion in silica and damaged $SiO_2$ have reported a broad range of activation energies, reflecting the diversity of trapping sites and material states (Griscom, 1983; Devine, 1985; Fink et al., 1995; Starukhina, 2006). Farrell et al. (2015, 2017) applied this idea to solar-wind hydroxylation by showing that a distribution of activation energies can strongly affect the retention of implanted H in lunar-like regolith. This range is important for the lunar surface because the regolith is not a single uniform material. Instead, the optically active surface contains a mixture of mineral grains, glasses, defects, amorphous rims, and space-weathered microstructures.

For this reason, the activation energy used in the model should not be interpreted as a unique mineralogical property of one site type. In the present work, we treat the activation-energy distribution as an effective description of the H population that contributes to the $M^3$-sensitive OH/$H_2O$ reservoir. This effective distribution accounts for the combined influence of implantation depth, trapping, detrapping, diffusion, and recombinative release within the upper grain rim.

The temperature dependence of this process naturally produces latitude and local-time trends. In warmer regions, especially near local noon, H atoms diffuse more rapidly and are more likely to reach the surface or recombine and release as $H_2$. In colder regions, diffusion is slower, leading to greater retention of implanted H and a stronger 3-micron absorption signature. Thus, the Starukhina-style implantation-diffusion pathway predicted both a latitudinal gradient and a diurnal variation in retained OH/$H_2O$ before the widespread lunar 3-micron feature was reported in 2009.

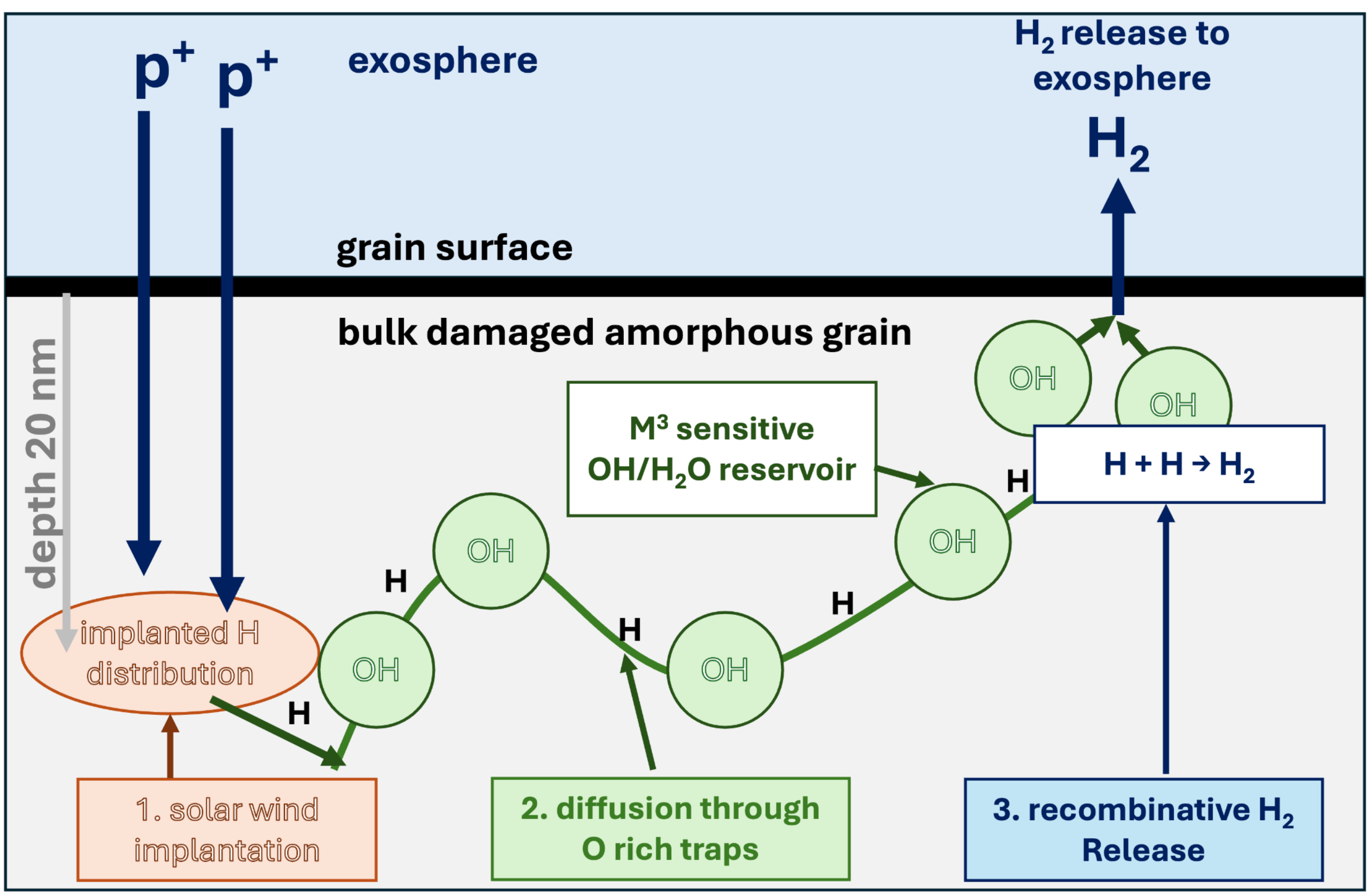

**Figure 1. Solar wind implantation, H diffusion, and $H_2$ release pathway.** Schematic illustration of the model pathway in which solar wind protons are implanted into the upper damaged rim of lunar regolith grains, become mobile H, occupy metastable O-H trapping sites, and are eventually lost through thermally controlled diffusion and recombinative $H_2$ release. The activation-energy distribution represents an effective description of the $M^3$-sensitive retained H/OH reservoir rather than a unique mineralogical activation-energy spectrum.

Laboratory and modeling studies provide additional support for key parts of this pathway, while also showing that multiple reaction channels may operate at the lunar surface. Several ion-irradiation experiments of lunar analogs and Apollo samples have shown that proton, deuterium, or molecular-hydrogen irradiation can produce OH-bearing spectral features near 3 microns, although interpretation can be complicated by terrestrial water contamination and sample preparation effects (Zeller et al., 1966; Burke et al., 2011; Ichimura et al., 2012; McLain et al., 2021; Yeo et al., 2025). More recent experimental work has improved sample preparation and bakeout procedures, strengthening the case that implanted hydrogen can form OH in lunar materials. Other experiments have shown that implanted deuterium can recombine and desorb as $D_2$, analogous to $H_2$ release following solar wind H implantation (Crandall et al., 2019; Zhu et al., 2019). Jones et al. (2018, 2024) modeled related solar-wind-driven lunar water production pathways, including OH recombination and meteoroid-impact contributions. Those studies are complementary to the present approach, but the dominant pathway followed here is the Starukhina implantation-diffusion pathway in which mobile implanted H is retained as an OH-like population and is released primarily as $H_2$ through recombinative desorption.

Tucker et al. (2019, 2021) incorporated this physical picture into a global Monte Carlo surface diffusion-exosphere model. In that model, solar wind implanted H atoms are assigned activation energies from a distribution, diffuse within the surface, and are released as $H_2$ through recombinative desorption. Tucker et al. (2019, 2021) showed that a Gaussian activation-energy distribution centered near 0.5 eV with a width of approximately 0.08 eV could reproduce both the $M^3$-inferred surface $OH/H_2O$ abundance and the Lunar Reconnaissance Orbiter Lyman Alpha Mapping Project (LRO-LAMP) constraints on the lunar $H_2$ exosphere. The present work extends that approach by comparing the model directly to the Li et al. (2023) phase- and local-time-resolved $M^3$ data set.

## 3. $M^3$ Data Set and Phase/Local-Time Sampling

The Chandrayaan-1 spacecraft was launched in 2008 and carried the Moon Mineralogy Mapper, an imaging spectrometer that measured reflected solar radiation and thermal emission from the lunar surface from 0.43 to 3.0 microns with 10 nm spectral sampling (Green et al., 2011). $M^3$ observations revealed a near 3-micron absorption feature across sunlit regions of the Moon, interpreted as an $OH/H_2O$-related spectral signature in the upper optical surface of the regolith (Pieters et al., 2009).

Several independent analyses of $M^3$ spectra have investigated the spatial distribution, abundance, and time-of-day dependence of this feature. These studies generally agree that $OH/H_2O$-related absorption is widespread, but they differ in the magnitude of the inferred abundance and the strength of the diurnal variation. These differences arise primarily from the difficulty of removing the thermal emission component near 3 microns and from different approaches to spectral correction and calibration. Bandfield et al. (2018), for example, emphasized that time-of-day and latitude trends can be near the uncertainty limits of the thermal correction, whereas Woehler et al. (2017), Li and Milliken (2017), and Li et al. (2023) reported clearer diurnal trends.

For this study, we use the high-level $M^3$ data set produced by Li et al. (2023). That data set was designed to examine the inferred $OH/H_2O$ abundance during intervals when the Moon passed through Earth's magnetosphere. The measurements are binned by broad absolute latitude ranges, local time, and lunar phase. The principal latitude bins are low latitude, 0-30 degrees; mid latitude, 30-60 degrees; and high latitude, 60-75 degrees, in both hemispheres. The data set also separates observations according to lunar phase, allowing the measurements to be related to the solar wind, magnetosheath, and magnetotail plasma environments. Previous low-latitude $M^3$ comparisons inside and outside the magnetotail also suggested changes in the 3-micron feature, although those measurements did not fully sample the diurnal structure emphasized here (Cho et al., 2018).

The source flux used in the model is scaled using a representative phase-dependent ion flux profile derived from THEMIS–ARTEMIS observations of the lunar plasma environment, as described by Poppe et al. (2018) and Li et al. (2023). In the present comparison, lunar phase is expressed relative to magnetotail center, with the central tail near 0 degrees, the dusk-side sheath before tail entry, and the dawn-side sheath after tail exit. Figure 2 summarizes the representative source-flux profile used in the model and the $M^3$ sampling geometry. The upper panel shows the phase-dependent ion flux used to scale the incident H source in the model, while the lower panel shows the $M^3$ comparison targets in phase and local time for the latitude bins used in the analysis.

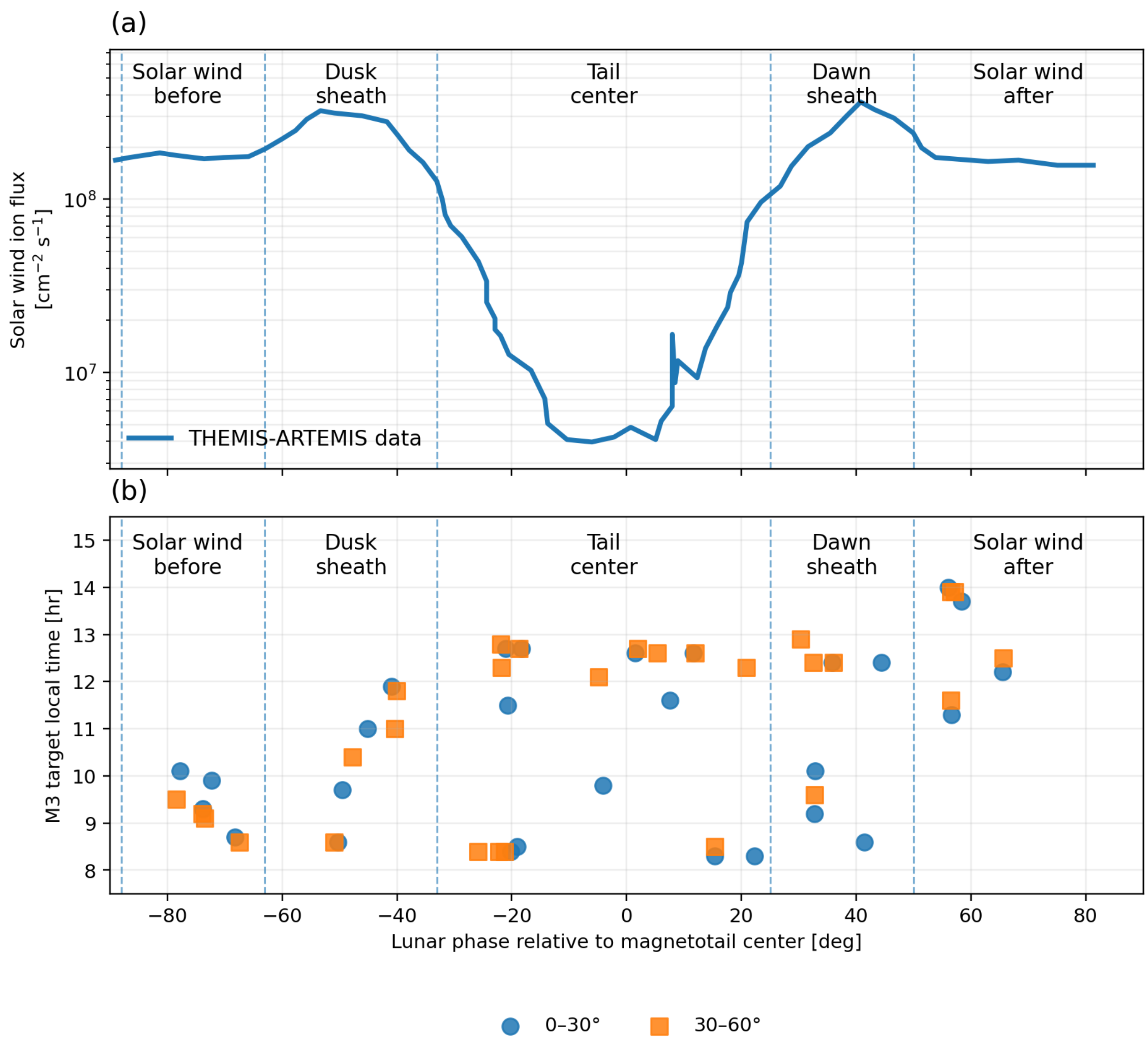

**Figure 2. Plasma source history and M³ comparison geometry.** (a) Representative phase-dependent THEMIS-ARTEMIS ion flux used to scale the incident implanted-H source in the model. Lunar phase is shown relative to magnetotail center. (b) Low- and mid-latitude M³ comparison targets plotted by lunar phase and local time. These sampling windows define the model-data comparison bins and show why the comparison must preserve phase, local time, and latitude rather than use a single averaged diurnal curve.

This phase/local-time representation is important because the M³ comparison is not simply a local-time average. Local time, lunar phase, plasma environment, and source history are coupled as discussed in Tucker et al. (2019, 2021). A given M³ point samples a surface location with a particular thermal history and a particular recent history of solar wind exposure, sheath enhancement, or magnetotail depletion. Therefore, as discussed below, the model is sampled at the corresponding phase and local-time bins rather than compared only to an idealized noon-normalized diurnal curve.

In the main model-data comparison, we focus on the low- and mid-latitude bins. These bins provide the cleanest test of thermally controlled diffusion because they are less affected by long-term cold-region accumulation and saturation effects. The high-latitude measurements are retained as a diagnostic comparison but are not used as the primary constraint on the activation-energy distribution. This choice follows the model behavior found in Tucker et al. (2019, 2021), where diffusion alone does not bring the coldest polar regions to equilibrium over short timescales without additional loss, saturation, or gardening processes.

## 4. Surface Diffusion-Exosphere Model

We use the global surface diffusion-exosphere Monte Carlo model developed in Tucker et al. (2019, 2021) as the baseline framework. The model represents the lunar surface as a spherical lower boundary divided into surface elements that track implanted H, retained near-surface H/OH populations, recombinative $H_2$ production, and $H_2$ transport in the exosphere. Solar wind protons are implanted into the dayside surface with a flux that depends on solar zenith angle and lunar phase. The local dayside source is weighted by the cosine of the solar zenith angle and set to zero on the nightside, consistent with the surface illumination and temperature treatment used in previous lunar exosphere modeling (Hurley et al., 2015; Tucker et al., 2019, 2021). The incident flux is scaled by the representative phase-dependent THEMIS-ARTEMIS ion flux profile shown in Figure 2. This profile is used as an average phase-dependent plasma forcing, and not as an event-specific reconstruction of the instantaneous proton flux during each $M^3$ observation. Following Chandrayaan-1/ Sub-keV Atom Reflecting Analyzer energetic-neutral-atom measurements showing that up to approximately 20% of incident solar-wind protons can be reflected from the lunar regolith as neutral H, and later 3D regolith simulations of solar-wind reflection, the model treats 80% of the phase-dependent proton flux as the implanting source and the remaining 20% as backscattered or otherwise not retained (Wieser et al., 2009; Futaana et al., 2012; Szabo et al., 2023).

In the Tucker et al. (2019, 2021) framework, implanted H is followed using a particle-based Monte Carlo treatment. Individual implanted H particles are assigned implantation depths and activation energies by random sampling from the prescribed depth and activation-energy distributions, and their retention or loss is evaluated according to the local surface temperature and thermally activated diffusion timescale. The activation-energy distribution used by Tucker et al. (2019, 2021) follows the statistical treatment discussed by Farrell et al. (2017), which was based in part on the experimental hydrogen-implantation and diffusion results of Fink et al. (1995). Those earlier model results were then compared with global $M^3$ and LRO-LAMP constraints. In the present work, we retain the same implantation, diffusion, and release physics, but treat the diffusion activation-energy distribution parameters as effective quantities constrained by the phase- and local-time-resolved $M^3$ measurements of Li et al. (2023). We compute the retained surface abundance using the weighted activation-energy-bin representation. In this approach, the continuous activation-energy distribution is divided into fixed bins with prescribed probability weights.

The implantation depth distribution is represented by a Gaussian centered near 20 nm, corresponding to the damaged amorphous grain rim where solar wind H is expected to be deposited. The underlying activation-energy distribution is also represented by a Gaussian,

$$P(E) \propto \exp[-(E - E_c)^2/(2E_w^2)], \quad \text{Eq. 2}$$

where $E$ is the activation energy, $E_c$ is the center of the effective activation-energy distribution, and $E_w$ is its width. The activation-energy distribution in Eq. 2 is truncated at an upper bound $E_{max}$. The diffusive lifetime of implanted H then depends on activation energy, implantation depth, and local surface temperature through the Arrhenius diffusion coefficient in Eq. 1. In this work, $D_0$ is kept fixed at the value used in Tucker et al. (2021), while $E_c$ and $E_w$ are varied to determine which effective distribution best matches the $M^3$ observations. As mentioned above, earlier work used $E_c$ and $E_w$ values motivated by prior studies of H diffusion and retention (Fink et al., 1995; Farrell et al., 2017). Here, we vary these parameters to identify the effective activation-energy distribution region that best reproduces the phase- and local-time-resolved $M^3$ trends in the detailed data set of Li et al. (2023).

The weighted activation-energy-bin representation approximates the continuous activation-energy distribution in Eq. 2 using a finite set of discrete energy bins. This discretization must be fine enough to resolve the low-energy portion of the distribution, which controls rapid thermal loss, and the higher-energy tail, which controls longer-lived retention. We tested the sensitivity of the energy-distribution-weighted retained abundance to the number of activation-energy bins and used 15 bins for the final refined-grid calculations. Coarser discretizations provided useful numerical checks, but the 15-bin representation more smoothly resolves the effective activation-energy distribution while remaining computationally practical for multi-lunation global simulations. Too few bins can underresolve the low- and high-energy tails, altering the balance between rapidly lost and longer-lived retained H populations.

For each surface element, the weighted-bin calculation tracks the contribution of the discretized activation-energy distribution to the retained H/OH inventory and computes an energy-distribution-weighted abundance, denoted $ppm_{\mathrm{Ebin}}$. This field reduces particle-counting noise in sparsely sampled phase, local-time, and latitude bins and provides the primary model quantity used in the $M^3$ comparison. The particle-based Monte Carlo abundance is retained as an independent model diagnostic and is compared with $ppm_{\mathrm{Ebin}}$ to verify that the weighted-bin representation preserves the large-scale behavior of the particle-based model. Figure 3 illustrates the numerical distinction between stochastic Monte Carlo sampling of activation energies and the weighted activation-energy-bin representation used to compute $ppm_{\mathrm{Ebin}}$.

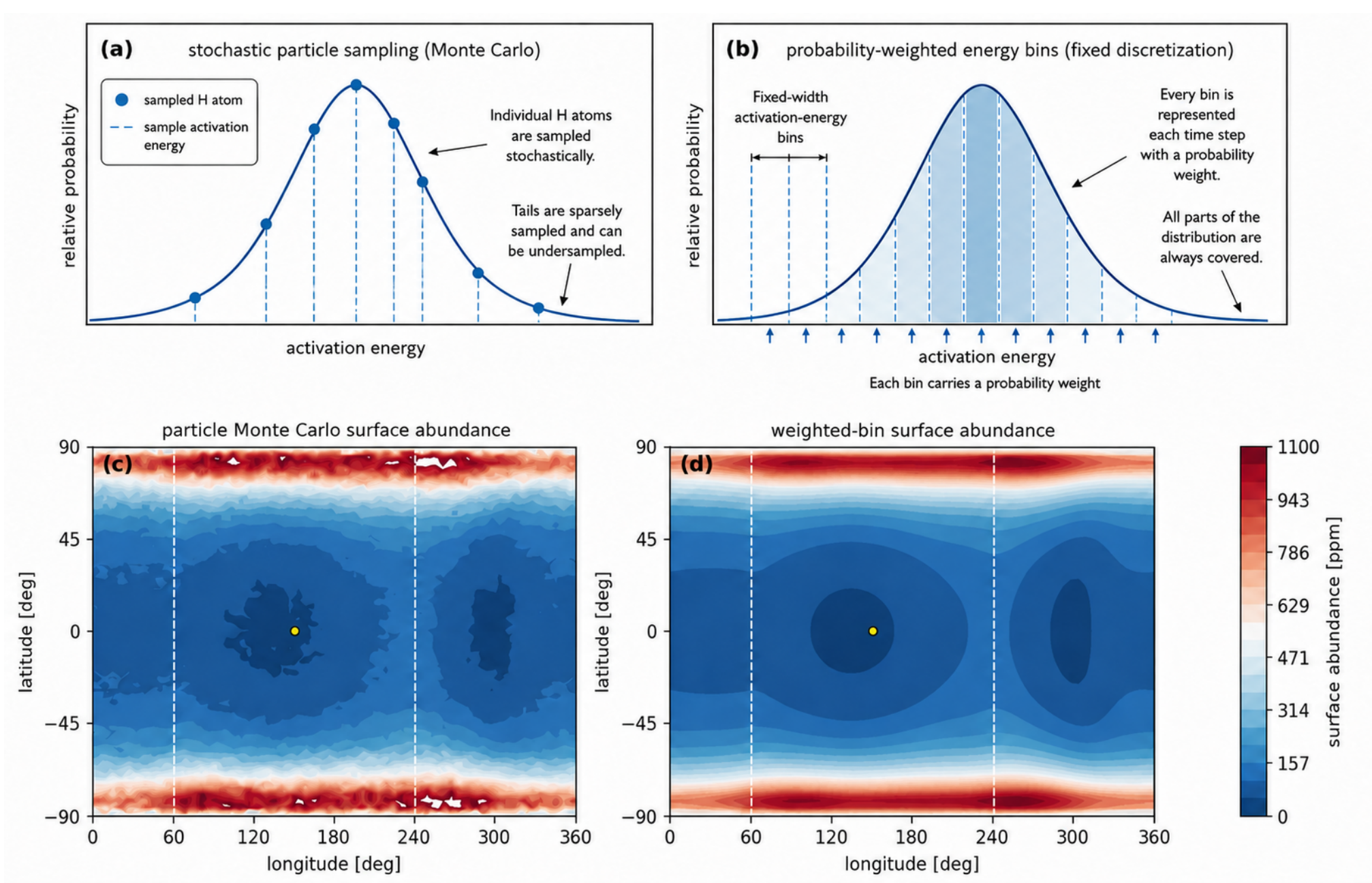

**Figure 3. Monte Carlo and weighted-bin representations of the activation-energy distribution.** (a) The particle Monte Carlo treatment assigns activation energies to implanted H atoms by stochastic sampling, which can sparsely represent low-probability portions of the distribution in low-flux surface elements. (b) The weighted activation-energy bin representation discretizes the same distribution into fixed probability-weighted bins, ensuring that all energy intervals contribute at each time step. (c,d) Example surface abundance maps from the same model snapshot show that the weighted-bin field preserves the large-scale structure of the particle Monte Carlo field while reducing sampling noise. The maps use the same abundance scale and illumination geometry; the yellow circle marks the subsolar point, and white dashed lines mark the terminators.

The use of an activation-energy distribution is central to the physical interpretation of the model. A single activation energy would imply a uniform trapping environment, whereas the lunar optical surface contains a range of defect sites, amorphous rims, grain coatings, local bonding environments, and mineralogical contexts (e.g., Farrell et al., 2015, 2017). The fitted $E_c$ and $E_w$ should therefore be interpreted as effective parameters describing the subset of implanted H that contributes to the $M^3$-sensitive reservoir and its loss to $H_2$. They should not be interpreted as a unique laboratory activation-energy for one mineral or one defect type.

The model follows the Starukhina-style pathway in which implanted H diffuses through oxygen-rich trapping sites and can form metastable OH bonds. These retained H/OH populations are used as the model counterpart to the $M^3$-inferred OH/$H_2O$ abundance. When two mobile H atoms interact near the surface, they can recombine and desorb as $H_2$. The released $H_2$ molecules are then tracked in the exosphere using computational particles in a spherical grid, following the H-to-$H_2$ surface-release pathway previously compared with LRO-LAMP constraints on the lunar $H_2$ exosphere (Cook et al., 2013; Hurley et al., 2017; Tucker et al., 2019).

For the adopted implantation scale and diffusion prefactor, the populated portion of the preferred activation-energy distribution is mobile over low- and mid-latitude daytime conditions. At the flux-weighted dayside temperature of $\sim$ 365–370 K, H atoms with $E \sim 0.5$ eV diffuse on approximately hour timescales, while atoms with $E \sim$ 0.7–0.8 eV diffuse over weeks to about a year. For the preferred distribution, approximately 99.9% of implanted atoms are assigned activation energies $\leq 0.8$ eV, so the 20-lunation simulation is sufficient for the low- and mid-latitude retained H distributions to approach a quasi-repeating state. This should not be interpreted as full global equilibrium, because colder high-latitude reservoirs can continue evolving over longer timescales. In the present work, the model output is sampled over the final lunations of the run and compared with $M^3$ points using the corresponding lunar phase, local time, and latitude bins. Model outputs include the particle-based retained surface abundance, the energy-distribution-weighted retained abundance $ppm_{\mathrm{Ebin}}$, and the $H_2$ exosphere density. For the primary $M^3$ comparison, we use $ppm_{\mathrm{Ebin}}$, the energy-distribution-weighted retained H/OH-like surface abundance expressed in ppm-equivalent units. This conversion assumes a surface-layer bulk mass density of 1500 kg $m^{-3}$ while the Monte Carlo surface abundance is used to demonstrate consistency with the particle-based model framework.

## 5. Model-Data Comparison and Activation-Energy Parameter Search

The primary goal of the model-data comparison is to determine whether the Li et al. (2023) $M^3$ phase/local-time trends can be reproduced by solar wind implantation and thermally controlled H diffusion, and to identify the effective activation-energy distribution required to do so. To constrain the effective activation-energy distribution, we compare model outputs directly to the $M^3$ data across the sampled phase, local-time, and latitude bins.

For each model case, the Gaussian activation-energy distribution is specified by $E_c$, $E_w$ and $E_{max}$. The model is run through repeated lunations using the phase-dependent THEMIS-ARTEMIS ion flux profile derived from ARTEMIS observations following Li et al. (2023). The $M^3$ comparison is then performed by sampling the model at the same lunar phase and local-time bins as the observations. This avoids mixing observations from different source environments and preserves the coupling among source flux, local time, temperature, and recent exposure history.

The main parameter search varies $E_c$ and $E_w$ while holding the diffusion prefactor and other model assumptions fixed. The parameter search uses the low- and mid-latitude bins as the primary fitting constraint. For model-observation comparison, we calculate the model-data residual,

$$r_i = M_i - O_i, \qquad \text{Eq. 3}$$

where $M_i$ is the model abundance sampled at the i-th $M^3$ comparison bin and $O_i$ is the corresponding $M^3$ observed abundance. The residual defined in Eq. 3 is used to calculate the root mean square error (RMSE)

$$RMSE = \left(\frac{1}{N}\sum_i r_i^2\right)^{1/2} \quad \text{Eq. 4}$$

and the mean bias

$$bias = \frac{1}{N}\sum_i r_i. \quad \text{Eq. 5}$$

Here, an individual residual is the signed model–observation difference for one comparison point. The mean bias is the average signed residual over a selected group of points. Negative bias indicates model underprediction, whereas positive bias indicates model overprediction. Mean absolute residual and maximum absolute residual describe residual magnitudes without regard to sign. We identify the preferred parameter region by minimizing RMSE (Eq. 4), while using the mean bias (Eq. 5) to evaluate whether a model case systematically overpredicts or underpredicts the $M^3$ abundance.

The resulting parameter search is interpreted as a constraint on an effective activation-energy response region rather than a unique best-fit pair of parameters. Because changes in $E_c$ and $E_w$ can compensate each other, the fit may produce a shallow valley in parameter space. In this case, a range of distributions can yield similar $M^3$ comparisons. This degeneracy is physically reasonable because the $M^3$-sensitive reservoir likely reflects a combination of implantation depth, trapping site diversity, temperature history, and release pathways.

## 6. Results

### 6.1. Activation-energy parameter search

We first evaluated the model over a grid of Gaussian activation-energy distributions and compared each case directly to the low- and mid-latitude $M^3$ comparison points. The comparison was made at the corresponding lunar phase, local time, and latitude bins. This sampling preserves the coupling between surface temperature, recent source history, and plasma environment across the magnetotail, magnetosheath, and solar wind intervals.

Figure 4 shows the resulting root-mean-square error and mean bias for the low- and mid-latitude comparison. The preferred region of parameter space is centered near $E_c$ = 0.520 eV and $E_w$ = 0.090 eV. For the combined low- and mid-latitude comparison, the preferred parameters give an RMSE of 27.2 ppm and a mean bias of -3.6 ppm, indicating that the model reproduces the average $M^3$ abundance closely while slightly underpredicting it overall. The neighboring case $E_c$ = 0.525 eV and $E_w$ = 0.085 eV gives a similar RMSE of 27.7 ppm and a bias of -6.3 ppm.

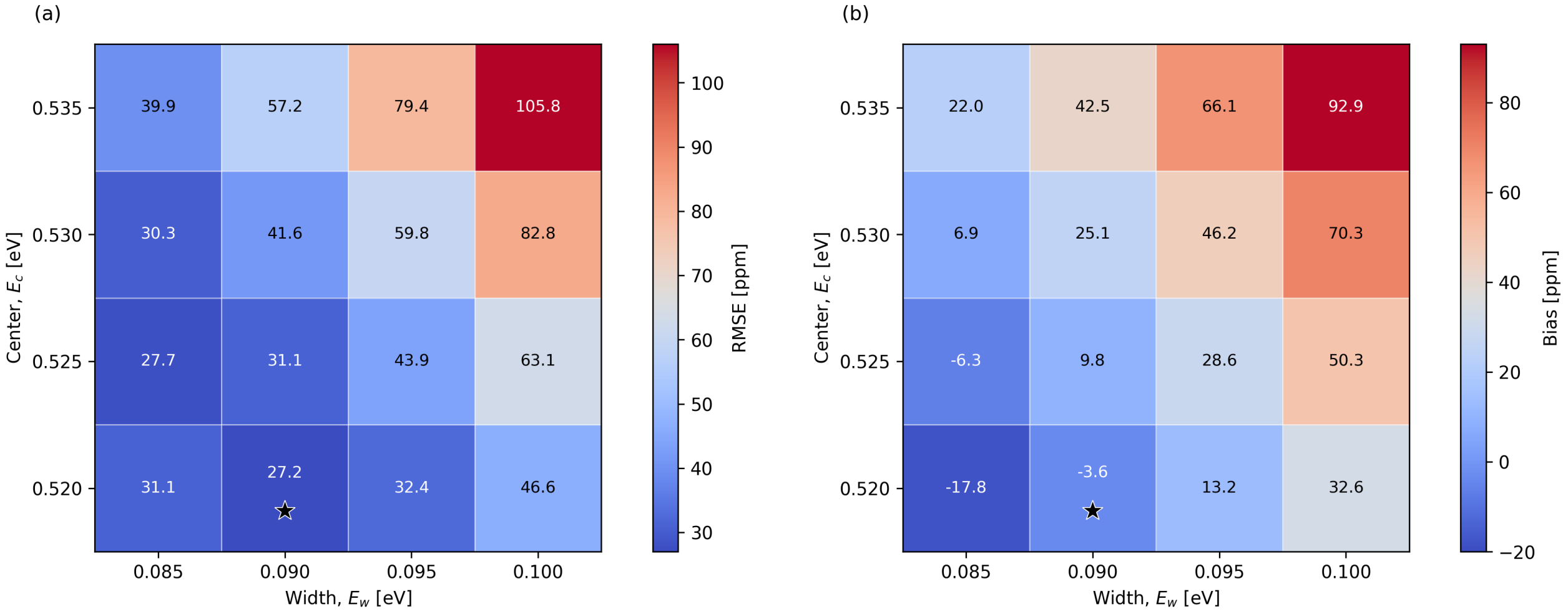

**Figure 4. Activation-energy parameter search.** Low- and mid-latitude model-data comparison metrics for the Gaussian effective activation-energy distribution. (a) RMSE (Eq. 4) between model and $M^3$ inferred surface abundance. (b) Mean bias (Eq. 5) model minus $M^3$ abundance. The star marks the preferred grid case which has the lowest RMSE and a mean bias close to zero. The shallow minimum near $E_c$ = 0.520 eV and $E_w$ = 0.090 eV indicates a preferred effective activation-energy response region for the $M^3$-sensitive reservoir, not a unique mineralogical activation energy.

The best-fit region is a shallow minimum in effective activation-energy parameter space rather than a unique best-fit activation energy. Nearby distributions produce comparable low- and mid-latitude fits, indicating that the $M^3$ comparison constrains a preferred response region rather than a single unique parameter pair. For example, the cases $E_c$ = 0.525 eV, $E_w$ = 0.085 eV and $E_c$ = 0.530 eV, $E_w$ = 0.085 eV bracket the preferred solution, with the former tending to underpredict the mean abundance and the latter tending to overpredict it. This pattern shows that the $M^3$ comparison is sensitive to both the center and width of the effective activation-energy distribution.

## 6.2. Best-fit low- and mid-latitude comparison

Figure 5 compares the preferred model case, $E_c$ = 0.520 eV and $E_w$ = 0.090 eV, directly with the $M^3$ derived surface abundances for the low- and mid-latitude bins. The model reproduces the overall range of the $M^3$ points and captures much of the separation between depleted and enhanced local-time conditions. The agreement is strongest in the sense of reproducing the broad amplitude of the observed low- and mid-latitude variations, rather than matching every phase/local-time point exactly.

The vertical bars in Figure 5 are not measurement uncertainties. In Figure 5(a, b) they show the range of model abundances sampled across the latitude interval represented by each $M^3$ comparison point. This range is useful because each low- or mid-latitude $M^3$ bin corresponds to a finite latitude region rather than a single model cell.

The bars therefore indicate model sampling variability within the finite M³ comparison bin and should not be interpreted as formal observational uncertainty.

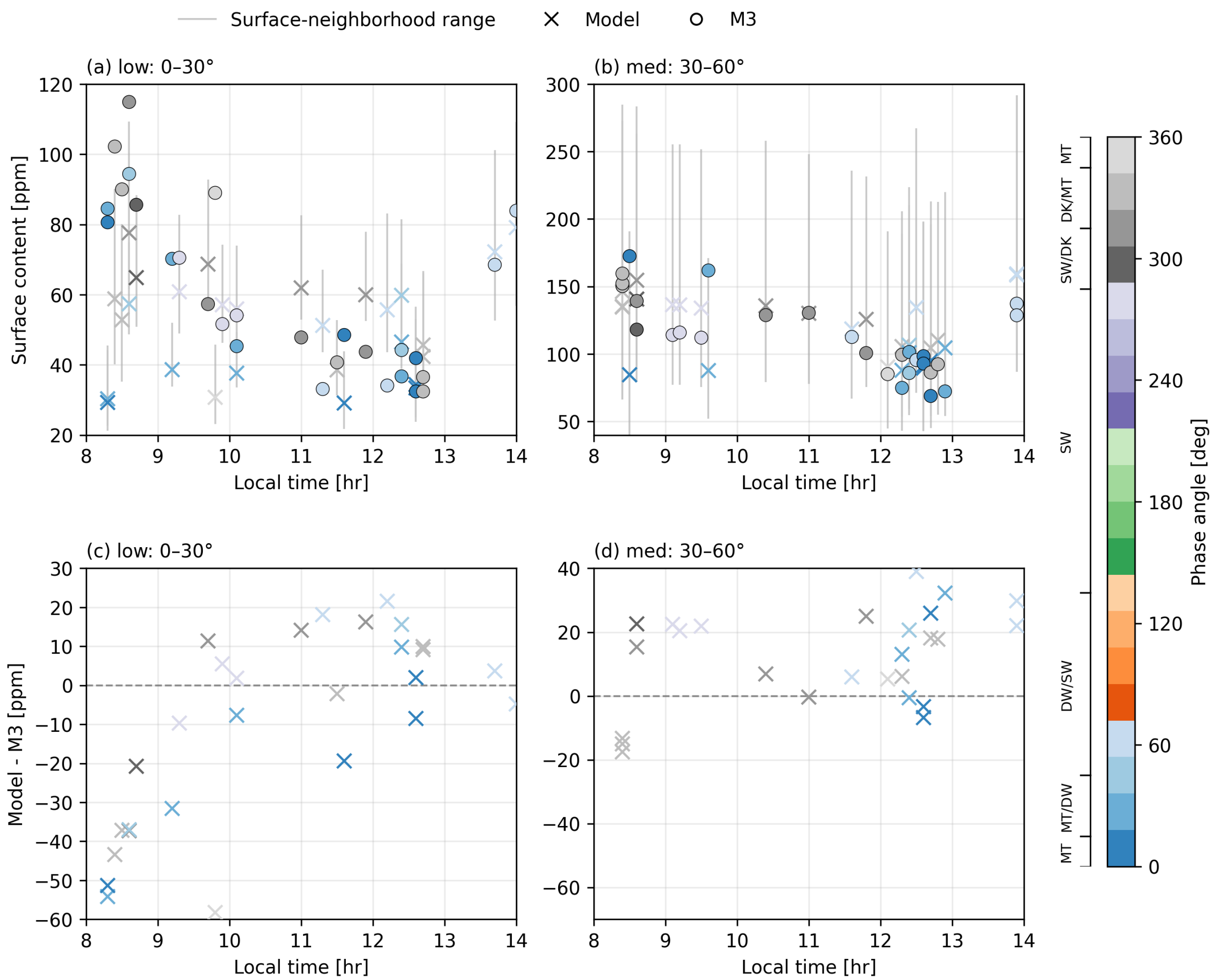

**Figure 5. Phase-averaged M³ measurement versus model comparison for $E_c$ = 0.520 eV and $E_w$ = 0.090 eV.** M³-inferred surface abundances are compared with model abundances sampled at the corresponding phase, local-time, and latitude bins. Panels (a) and (b) show M³ and model surface abundance as a function of local time for the low-latitude 0–30 degree and mid-latitude 30–60 degree bins, respectively. Model values are shown with X symbols, and M³ values are shown with circles. Gray vertical bars in Panels (a) and (b) show the range of model abundances sampled within the corresponding latitude and local-time bin. Panels (c) and (d) show signed residuals, defined as model minus M³ abundance. Points are colored by lunar phase angle, with magnetospheric-sector labels shown next to the color bar.

The low-latitude comparison shows a modest but systematic underprediction in the preferred case, while the mid-latitude comparison shows a smaller positive bias. For the same case, the low-latitude RMSE is approximately 26 ppm with a mean bias of about −10 ppm, whereas the mid-latitude RMSE is approximately 29 ppm with a mean bias of about +5 ppm. When the two latitude ranges are combined, these latitude-dependent offsets partially compensate, producing a combined low- and mid-latitude RMSE of approximately 27 ppm. The low-latitude underprediction is most apparent at selected morning local times, including points near 8–9 hr local time, where several model values fall below the corresponding $M^3$ abundances. In this comparison, lunar phase identifies the Moon's plasma-environment position relative to magnetotail center, whereas local time identifies the illumination state of the observed surface bin. Thus, measurements near magnetotail center can still sample morning or afternoon local-time bins depending on the observed longitude and $M^3$ spatial coverage.

Despite these residuals, the model captures the first-order amplitude and local-time dependence of the $M^3$ variations. The lowest $M^3$-inferred abundances generally occur near local noon, when surface temperatures are highest and diffusive loss is most efficient, whereas higher abundances occur closer to the cooler morning and afternoon local times. This local-time structure is consistent with thermally controlled retention and loss of implanted H. The remaining coherent residuals may reflect a combination of $M^3$ averaging choices, finite phase/local-time sampling, latitude-dependent surface response, and additional surface or plasma processes not represented in the present diffusion-focused model.

Table 1. Summary statistics comparing modeled and observed $OH/H_2O$ abundance trends.

| Group | N | Mean bias (ppm) | RMSE (ppm) | Mean absolute residual (ppm) | Max absolute residual (ppm) |
|---|---|---|---|---|---|
| All points | 56 | -2.37 | 27.31 | 20.57 | 87.87 |
| Low lat., 0–30° | 28 | -10.16 | 26.06 | 20.08 | 58.27 |
| Mid lat., 30–60° | 28 | 5.42 | 28.51 | 21.06 | 87.87 |
| MT | 8 | -7.89 | 24.00 | 16.19 | 58.27 |
| MT/DW | 13 | -19.47 | 42.24 | 33.52 | 87.87 |
| DW/SW | 8 | 16.93 | 21.68 | 18.12 | 39.01 |
| SW/DK | 11 | 4.63 | 19.05 | 16.95 | 37.30 |
| DK/MT | 14 | -1.90 | 20.24 | 16.53 | 43.40 |

*Statistics are reported for all points, latitude subsets, and magnetospheric/local-time sectors. The mean bias is the average signed residual, Eq. 5. The mean absolute residual and maximum absolute residual describe residual magnitudes. Negative mean bias indicates model underprediction, whereas positive mean bias indicates model overprediction.*

Table 1 quantifies the visual pattern in Figure 5. The preferred parameter set underpredicts the low-latitude bins on average and slightly overpredicts the mid-latitude bins. When grouped by plasma/local-time sector, the largest residuals occur in the MT/DW transition group, which has the largest RMSE and maximum absolute residual. The lower RMSE values for the DW/SW, SW/DK, and DK/MT groups indicate that the largest discrepancies are concentrated in mixed plasma-transition intervals rather than uniformly distributed across all phase bins.

### 6.3. Phase and local-time structure of the residuals

Figure 6 shows the same preferred case as a set of local-time profiles sampled at the nearest 30-degree model phase bins. The $M^3$ points are overplotted in the corresponding local-time and latitude bins. The shaded envelopes in Figure 6 show the model min–max range across each sampled latitude interval, with the low-latitude 0–30° range shown in blue and the mid-latitude 30–60° range shown in orange. The model reproduces the main local-time behavior expected from thermally controlled diffusion. That is, the retained surface abundance is generally lower near the warmest local times and higher toward cooler morning or afternoon conditions. This pattern is present in both the low- and mid-latitude profiles, although the amplitude and scatter differ by phase bin. The result supports the central interpretation that a solar wind implantation and thermal-diffusion model can reproduce the dominant low- and mid-latitude $M^3$ structure without requiring each individual $M^3$ point to be fit independently.

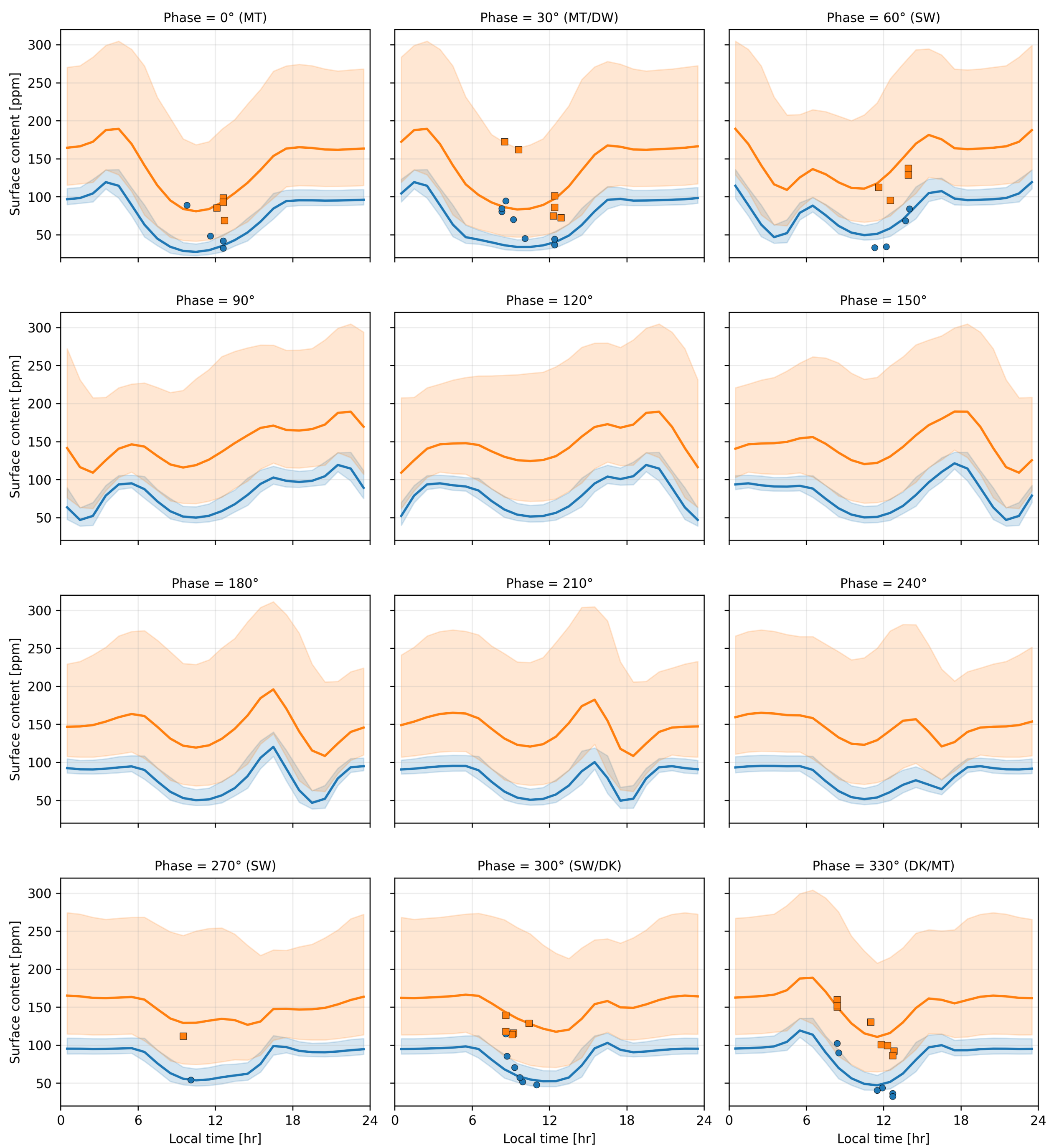

**Figure 6. Phase- and local-time-resolved comparison for the preferred case.** Model local-time profiles and M[3] comparison points for $E_c$ = 0.520 eV and $E_w$ = 0.090 eV, grouped by nearest 30-degree model phase bin. Blue shading shows the model abundance range across the low-latitude 0–30° sampling interval, and orange shading shows the corresponding range across the mid-latitude 30–60° interval. Solid curves show the mean model abundance within each latitude interval. M[3] points are plotted in the corresponding local-time and latitude bins. Phase labels identify the plasma regions sampled by the $M^3$ points: MT, magnetotail; DW, dawn sheath; SW, solar wind; DK, dusk sheath. Mixed labels indicate phase bins containing $M^3$ points from adjacent plasma regions.

Several of the largest differences occur in phase windows near magnetotail and sheath transitions. In these regions, the $M^3$ measurements are assigned to the nearest 30-degree model phase bin, whereas the actual $M^3$ target phases span finite windows. The largest deviations occur in the transition-region panels, especially near the MT/DW and DK/MT phase bins and at morning local times. This pattern suggests that the strongest discrepancies are associated with plasma-transition sampling rather than with the central magnetotail or the local-noon thermal minimum.

## 7. Discussion

### 7.1. Effective activation-energy distribution and model interpretation

The preferred values found here, $E_c = 0.520$ eV and $E_w = 0.090$ eV, are close to the activation-energy distributions used in earlier lunar H-retention models, including the characteristic ~ 0.5 eV values in Farrell et al. (2017) and the Gaussian distribution centered near 0.5 eV with width 0.08 eV used by Tucker et al. (2019, 2021). Thus, the present $M^3$ comparison independently constrains nearly the same effective activation-energy response range using phase- and local-time-resolved observations. This result is also broadly consistent with the range of activation-energy behavior expected for implanted H in damaged, oxygen-bearing, silica-bearing, and amorphous surface materials, where trapping-site diversity and material state strongly influence the apparent diffusion response (Griscom, 1983; Devine, 1985; Fink et al., 1995; Starukhina, 2006; Farrell et al., 2015; Morrissey et al., 2022). Laboratory lunar-irradiation and thermal-release studies provide complementary constraints, but they probe specific exposure, heating, and sample-preparation windows rather than the full lunar diurnal/lunation response sampled by $M^3$ (McLain et al., 2021; Yeo et al., 2025; Crandall et al., 2019; Zhu et al., 2019). That is, laboratory diffusion studies constrain thermally activated transport over controlled material, temperature, and time windows, while lunar irradiation and thermal-release experiments constrain the subset of implanted H or D that is mobile or releasable over laboratory exposure and heating schedules. By contrast, $M^3$ samples the optically active reservoir that survives and evolves across lunar local-time and phase histories (Li and Milliken, 2017; Woehler et al., 2017; Bandfield et al., 2018; Li et al., 2023). Low-activation-energy sites may release H rapidly under warm daytime conditions, whereas high-activation-energy sites can remain effectively immobile over the same observing window. The $M^3$ comparison is therefore most sensitive to the intermediate population that can be retained during cooler portions of the diurnal cycle but depleted under warmer conditions.

The ability of this effective distribution to reproduce the first-order low- and mid-latitude $M^3$ trends supports the view that solar wind implantation followed by thermally controlled diffusion is a major contributor to the observed 3-micron variability. The model captures the broad reduction in retained abundance near warmer local times and the enhanced retention toward cooler morning and afternoon conditions. The updated low- and mid-latitude comparison is consistent with the interpretation that thermally controlled diffusion and $H_2$ release contribute substantially to the local-noon depletion, without requiring an additional magnetotail-specific production mechanism in the present model framework. This does not require that all $M^3$ inferred $OH/H_2O$ be newly produced during each lunation, nor does it exclude other pathways for $OH/H_2O$ formation or modification. Rather, it shows that the dominant low- and mid-latitude phase/local-time structure can be reproduced by a physically motivated implantation-diffusion pathway using an effective activation-energy distribution. This is important for predicting variations during extreme solar events when the proton flux can be more than an order of magnitude larger than nominal solar conditions (Farrell et al., 2012).

The present study differs from earlier work by using the phase- and local-time-resolved $M^3$ data to constrain the effective values of $E_c$ and $E_w$. The similarity between the preferred $M^3$-constrained distribution and earlier laboratory-motivated values supports the physical plausibility of the implantation-diffusion interpretation. Nevertheless, the fitted parameters should still be interpreted as effective parameters for the $M^3$-sensitive reservoir rather than as unique material properties.

### 7.2. Limitations and high-latitude diagnostic behavior

Although the preferred model reproduces the first-order $M^3$ trends, the residuals do not appear to be completely random. Some of the largest differences occur near magnetotail, sheath, and solar-wind transition intervals, where the $M^3$ measurements span finite phase windows, but the model comparison is assigned to discrete phase bins. In these regions, relatively small differences in source history, local time, or phase assignment can produce coherent residuals. This sensitivity is expected because the retained surface abundance depends not only on the instantaneous plasma source but also on the recent thermal and implantation history of the surface.

Additional physical processes may also contribute to the remaining residuals. The model emphasizes solar wind implantation, thermally activated diffusion, and recombinative $H_2$ release. Processes such as electron-stimulated chemistry or plasma-sheet interactions, sputtering and energetic neutral atom production, micrometeoroid gardening, impact-driven volatile release, and unresolved grain-scale or topographic temperature variations are not treated explicitly in the current comparison (Li et al., 2023; Futaana et al., 2012; Szabo et al., 2023; Benna et al., 2019; Szalay et al., 2019; Jones et al., 2018, 2024). These processes may be especially important in transition regions where the plasma environment changes rapidly or where $M^3$ sampling combines surfaces with different recent exposure histories. Because a single activation-energy distribution is applied globally, the fitted distribution should be interpreted as an effective low- and mid-latitude response rather than as a local material property. Local retention behavior may vary with composition, maturity, glass abundance, defect density, weathering state, unresolved temperature structure, and the presence of ice-bearing or strongly hydrated terrains. Furthermore, the present comparison uses an average THEMIS–ARTEMIS derived phase-dependent plasma forcing rather than the instantaneous plasma conditions at the time of each $M^3$ observation, which occurred before ARTEMIS was in lunar orbit. This limitation is most important near plasma-transition regions, where small differences in source history or phase binning can lead to larger model-data differences. Therefore, the residuals should be viewed as diagnostic of where additional surface or plasma processes may occur, not as evidence against the implantation-diffusion pathway.

Similar to the treatment in Tucker et al. (2019, 2021), high-latitude behavior is also treated as diagnostic rather than as a primary fitting constraint. Cold high-latitude regions retain implanted H more efficiently and can be more sensitive to saturation assumptions, unresolved temperature structure, regolith turnover, and the approach to quasi-steady state. Small changes in the assumed saturation limit, gardening rate, topographic temperature distribution, or model spin-up time can have a larger effect at high latitude than in the warmer low- and mid-latitude bins. For this reason, the low- and mid-latitude $M^3$ comparison provides the main constraint on the effective activation-energy distribution, while high-latitude comparisons are better suited for testing additional loss, saturation, and surface-evolution physics.

Taken together, the residual structure suggests that the present model captures the dominant thermal-diffusion control on the $M^3$-sensitive reservoir but does not yet describe all spatial, temporal, and plasma-environment-dependent variability. The effective activation-energy distribution should therefore be viewed as a compact parameterization of the reservoir sampled by $M^3$ over lunar diurnal and lunation timescales. During magnetosheath and magnetotail crossings, for example, the angular distribution of incident plasma and the resulting surface irradiation pattern may differ from the nominal solar-wind geometry assumed here. Plasma-environment-dependent irradiation maps would be a useful extension of the model, but are beyond the scope of the present comparison. Future work could test whether seasonal solar declination, signed-hemisphere comparisons, more detailed plasma-transition histories, or additional loss mechanisms improve the remaining residuals without changing the primary low- and mid-latitude constraint.

## 8. Conclusions

We compared a global solar wind implantation, surface diffusion, and $H_2$ exosphere model with phase- and local-time-resolved $M^3$ measurements of the lunar 3-micron OH/$H_2O$-related absorption. The model was sampled at the corresponding lunar phase, local time, and latitude bins, allowing the comparison to preserve the coupling among source flux, thermal history, and plasma environment. The low- and mid-latitude $M^3$ trends are reproduced to first order by a thermally controlled diffusion model using an effective Gaussian activation-energy distribution. The preferred region of parameter space is centered near $E_c$ = 0.52 eV with a width near $E_w$ = 0.09 eV. Neighboring parameter combinations provide similar fits, indicating that the result should be interpreted as a preferred effective response region rather than a unique best-fit activation-energy.

The fitted activation-energy distribution describes the $M^3$-sensitive retained H/OH reservoir active over lunar diurnal and lunation timescales. It should not be interpreted as a unique mineralogical activation-energy spectrum or as a single laboratory diffusion barrier. Instead, it parameterizes the combined effects of implantation depth, trapping-site heterogeneity, temperature history, diffusion, and recombinative loss for the portion of implanted H sampled by $M^3$. The remaining residuals are largest in selected phase and transition-region bins and likely reflect a combination of finite phase/local-time sampling, source-history effects, unresolved surface heterogeneity, and additional plasma or surface processes not included explicitly in the present diffusion-focused model. High-latitude behavior is retained as a diagnostic of saturation, cold-region retention, and additional loss physics, while the low- and mid-latitude bins provide the primary constraint on the effective activation-energy distribution.